\newcommand{\BF}[1]{\mbox{\boldmath $#1$}}

\def\dddot#1{{\hspace{3pt}\dot{\phantom #1}\hspace{-2.4pt}\ddot{\!#1}}}

\documentstyle[eqsecnum,aps]{revtex}

\def\aut#1{#1}

\def\cl{{\rm cl}}

\def\ins#1{}
\def\iem#1{{\em #1\/}}
\def\iems#1{}
\def\comment#1{}

 \newcommand{\tab}{_{t_a}^{t_b}}
\def\Det{{\rm  Det\,}}

\def\aut#1{#1}

\def\cm#1{}
\def\sbf#1{\mbox{\scriptsize{\bf #1}}}

 \def\lfrac#1#2{{{{#1}/{#2}}}}

\begin{document}
\setcounter{figure}{0}
\Roman{figure}

\title{{
Master Equation for Electromagnetic
Dissipation and Decoherence
of Density Matrix
}}
\author{Z. Haba%
 \thanks{On leave from Institute of Theoretical Physics, University of
Wroclaw, Poland; e-mail: zhab@ift.uni.wroc.pl}
 and H. Kleinert%
 \thanks{Email: kleinert@physik.fu-berlin.de 
URL:
http://www.physik.fu-berlin.de/\~{}kleinert \hfill
} }
\address{Institut f\"ur Theoretische Physik,\\
Freie Universit\"at Berlin, Arnimallee 14,
14195 Berlin, Germany}
\maketitle
\begin{abstract}
We set up a
forward--backward path integral for a point particle
 in a bath of photons
to derive a master equation for the density matrix
which describes electromagnetic dissipation and decoherence.
As an application, we recalculate the
Weisskopf-Wigner formula for the natural
line width of an atomic state
at zero temperature
and find, in addition, the temperature broadening
caused by the decoherence term.
\end{abstract}

%
\section{Introduction}
The time evolution
of
a quantum-mechanical density matrix
$ \rho ({\bf x}_{+a},{\bf x}_{-a};t_a)$
of a particle coupled to an external electromagnetic
vector potential ${\bf A}({\bf x},t)$ is determined by
a forward--backward path integral
  \cite{FV}
\begin{eqnarray}
&&\!\!\!\!\!\!\!\!\!( {\bf x}_{+b},t_b|{\bf x}_{+a},t_a)
( {\bf x}_{-b},t_b|{\bf x}_{-a},t_a)^*
\equiv
U({\bf x}_{+b},{\bf x}_{-b},t_b|
{\bf x}_{+a},{\bf x}_{-a},t_a
)\nonumber \\
&&=\int
{\cal D}{\bf x}_+
{\cal D}{\bf x}_-\,\exp\left\{ \frac{i}{\hbar }
\int\tab\left[
\frac{M}{2}\left(\dot {\bf x}_+^2-\dot {\bf x}_-^2\right)
-V({\bf x}_+)+V({\bf x}_-)
-\frac{e}{c}\dot {\bf x}_+{\bf A}({\bf x}_+,t)
+\frac{e}{c}\dot {\bf x}_-{\bf A}({\bf x}_-,t)
\right] \right\},
\label{@PRO}\end{eqnarray}
where
${\bf x}_+(t)$ and
${\bf x}_-(t)$ are two fluctuating paths
connecting the initial and final points
${\bf x}_{+a}$ and
${\bf x}_{+b}$, and
${\bf x}_{-a}$ and
${\bf x}_{-b}$, respectively.
In terms of this expression, the density matrix
$ \rho ({\bf x}_{+b},{\bf x}_{-b};t_b)$
at a time $t_b$ is found from  that at an earlier  time $t_a$
by
the integral
\begin{eqnarray}
 \rho ({\bf x}_{+b},{\bf x}_{-b};t_b)=
\int d{\bf x}_{+a}\,d {\bf x}_{-a}
 \,
U({\bf x}_{+b},{\bf x}_{-b},t_b|
{\bf x}_{+a},{\bf x}_{-a},t_a
)\rho ({\bf x}_{+a},{\bf x}_{-a};t_a).
\label{@locev}\end{eqnarray}
The vector potential
 ${\bf A}({\bf x},t)$  is a superposition of oscillators
${\bf X}_{\bf k}(t)
 $
of frequency
$ \Omega_{\bf k} =c|{\bf k}|$ in a volume $V$:
\begin{equation}
{\bf A}({\bf x},t)=   \sum _{\bf k}c_{\bf k}({\bf x})
{\bf X}_{\bf k}(t),~~~~c_{\bf k}= \frac
{e^{i{\bf k}{\bf x}}}{\sqrt{2  \Omega_{\bf k}V}},~~~~~
\sum _{\bf k}= \int \frac{d^3kV}{(2\pi)^3}.
\label{@}\end{equation}
These  oscillators
are  assumed to be
in equilibrium
at a finite temperature $T$,
 where we shall write their
 time-ordered
 correlation functions as
 $G^{ij}_{{\bf kk}'}(t,t')=\langle \hat T\hat X^i_{\bf k}(t),
\hat X^j_{-\bf k'}(t')\rangle
=
\delta ^{ij\,\rm tr}_{{\bf kk}'}G_{  \Omega _{\bf k}}(t,t') \equiv
(\delta ^{ij}-k^ik^j/{\bf k}^2)G_{  \Omega _{\bf k}}(t,t')$,
the transverse Kronecker symbol
resulting from the sum over the two
polarization vectors
$\sum _{h=\pm}
 {\epsilon}^i({\bf k},h)
 {\epsilon}^{j*}({\bf k},h)$
of
the vector potential ${\bf A}({\bf x},t)$.
For a single oscillator of frequency $ \Omega $, one has
for $t>t'$:
\begin{equation}
G_{ \Omega }(t,t')=\frac{1}{2}\left[  A_{ \Omega }(t,t')+
 C_{ \Omega }(t,t')\right] =\frac{\hbar }{2M \Omega }
\frac{\displaystyle\cosh\frac{ \Omega}{2}\left[\hbar  \beta -i(t-t')
\right]}
{\displaystyle\sinh\frac{\hbar  \Omega \beta }{2}}
,~~~t>t'
\label{@expo in}\end{equation}
which is the analytic continuation
of the periodic imaginary-time Green function
to $\tau =it$.
The decompostion into
${A}_ \Omega (t,t')$ and ${C}_ \Omega (t,t')$
distinguishes
real and imaginary parts, which are
commutator
and anticommutator
functions of the
 oscillator
at temperature $T$:
$
{C}_ \Omega (t,t')\equiv \langle [\hat { X}(t),\hat { X}(t')]\rangle_T
 $ and $
{A}_ \Omega (t,t')\equiv \langle [\hat {X}(t),\hat {X}(t')]\rangle_T
 $, respectively.
The thermal average of the probability (\ref{@PRO})
is then given by the forward--backward path integral
\begin{eqnarray}
\!\!\!\!\!\!\!\!\!\!\!\!\!\!\!\!\!\!\!\!
U({\bf x}_{+b},{\bf x}_{-b},t_b|
{\bf x}_{+a},{\bf x}_{-a},t_a
) &=&
  \int {\cal D}{\bf x}_+(t)\int {\cal D}{\bf x}_-(t)\,
  \nonumber \\
\!\!\!\!\!\!\!\!\!\!\!\!\!\!\!\!\!\!\!\!\! &\times&     \exp\left\{
 \frac{i}{\hbar}\int\tab dt\,
    \left[ \frac{M}{2}({\dot {\bf x}_+}^2-{\dot {\bf x}_-}^2)-
    (V({\bf x}_+)-V({\bf x}_-))\right] +\frac{i}{\hbar }
    {\cal A}^{\rm FV}[{\bf x}_+,{\bf x}_-]\right\}.
 \label{18.389n}
\end{eqnarray}
where $\exp\{i{\cal A}^{\rm FV}[{\bf x}_+,{\bf x}_-]/\hbar \}$ is the Feynman-Vernon
{\em influence functional\/}.
The inluence action
$i{\cal A}^{\rm FV}[{\bf x}_+,{\bf x}_-]$ is the sum
of a
dissipative and a fluctuating  part
${\cal A}^{\rm FV}_D[{\bf x}_+,{\bf x}_-]$ and
${\cal A}^{\rm FV}_F[{\bf x}_+,{\bf x}_-]$, respectively,
whose explicit forms are
\begin{eqnarray}
\!\hspace{-.9cm}
{\cal A}^{\rm FV}_D[{\bf x}_+,{\bf x}_-]
&=&
  \frac{ie^2}{2 \hbar c^2} \int dt \int dt'\,\Theta(t-t')  \Big[
\dot {\bf x}_+ {{\bf C}}_{\rm b}({\bf x}_+\,t,{\bf x}'_+\,t') \dot{\bf x}_+'
-\dot {\bf x}_+ {{\bf C}}_{\rm b}({\bf x}_+\,t,{\bf x}'_-\,t') \dot{\bf x}_-'
\nonumber \\
&&~~~~~~~~~~~~~~~~~~~~~~~~~~~~~~~~
-\dot {\bf x}_- {\bf C}_{\rm b}({\bf x}_-\,t,{\bf x}'_+\,t') \dot{\bf x}_+'
-\dot {\bf x}_- {\bf C}_{\rm b}({\bf x}_-\,t,{\bf x}'_-\,t') \dot{\bf x}_-'
  \Big] .
 \label{18.148bt}
\end{eqnarray}
and
\begin{eqnarray}
\!\hspace{-.9cm}
{\cal {\bf A}}^{\rm FV}_F[{\bf x}_+,{\bf x}_-]
&=&
  \frac{ie^2}{2\hbar c^2 } \int dt \int dt'\,\Theta(t-t')  \Big[
 \dot {\bf x}_+ {{\bf A}}_{\rm b}({\bf x}_+\,t,{\bf x}'_+\,t') \dot{\bf x}_+'
-\dot {\bf x}_+ {{\bf A}}_{\rm b}({\bf x}_+\,t,{\bf x}'_-\,t') \dot{\bf x}_-'
\nonumber \\
&&~~~~~~~~~~~~~~~~~~~~~~~~~~~~~~~~
-\dot {\bf x}_- {{\bf A}}_{\rm b}({\bf x}_-\,t,{\bf x}'_+\,t') \dot{\bf x}_+'
+\dot {\bf x}_- {{\bf A}}_{\rm b}({\bf x}_-\,t,{\bf x}'_-\,t') \dot{\bf x}_-'
  \Big] .
 \label{18.148btA}
\end{eqnarray}
where ${\bf x}_\pm$,
 ${\bf x}_\pm'$ are short for ${\bf x}_\pm (t),~{\bf x}_\pm (t') $, and
$
{{\bf C}}_{\rm b}({\bf x}_-\,t,{\bf x}'_-\,t'),
{{\bf A}}_{\rm b}({\bf x}_-\,t,{\bf x}'_-\,t') $ are
$3\times 3$ commutator and anticommutator
functions of the bath of photons.
They are
sums of correlation functions
over the bath of the oscillators of
 frequency $ \Omega _{\bf k}$, each contributing
with a weight
$c_{\bf k}({\bf x})
c_{-\bf k}({\bf x}')
=e^{i{\bf k}({\bf x}-{\bf x}') }/2 \Omega _{\bf k}V$.
 Thus we may write
\begin{eqnarray} \label{@}
\hspace{-.0cm}C^{ij}_{ \rm b}({\bf x}\,t,{\bf x}'\,t')&\!=\!\!&
  \sum _{\bf k}
c_{-\bf k}({\bf x})
c_{\bf k}({\bf x}')
\left\langle [\hat { X}^i_{-\bf k}(t),\hat { X}^j_{\bf k}(t')]\right\rangle_T~~=
 - i\hbar \int \frac{d\omega'd^3k}{(2\pi)^4}
\sigma_{\bf k} ( \omega' )\delta ^{ij\,\rm tr}_{{\bf kk}}e^{i{\bf k}({\bf x}\!-\!{\bf x}') }\sin \omega'(t-t'),
\label{18.386a}
 \\
\hspace{-.0cm}A^{ij}_{\rm b}({\bf x}\,t,{\bf x}'\,t' )&\!=\!\!&
 \sum _{\bf k}
c_{-\bf k}({\bf x})
c_{\bf k}({\bf x}')
\left\langle\left\{\hat { X}^i_{-\bf k}(t),\hat { X}^j_{\bf k}(t')\right\}\right\rangle_T\!=\!
  \hbar \int \frac{d\omega'd^3k}{(2\pi)^4}\sigma_{\bf k} ( \omega' )
\delta ^{ij\,\rm tr}_{{\bf kk}}
 \coth\frac{\hbar  \omega' }{2k_BT}~e^{i{\bf k}({\bf x}\!-\!{\bf x}') }\cos \omega'(t\!-\!t')
 ,   \label{18.386}
\end{eqnarray}
where $\sigma_{\bf k}(\omega') $
is the spectral density contributed by the oscillator of momentum ${\bf k}$:
\begin{equation}
\sigma_{\bf k}(\omega') \equiv  \frac{2\pi }{2\Omega_{\bf k}}
 [ \delta(\omega '-\Omega_{\bf k})-\delta(\omega' +\Omega_{\bf k})] .
  \label{18.387}
\end{equation}
At zero temperature, we recognize in (\ref{18.386a}) and (\ref{18.386})
twice the imaginary and real parts of the Feynman propagator
of a massless particle
for $t>t'$, which in four-vector notation
with $k=( \omega /c,{\bf k})$ and $x=(ct,{\bf x})$ reads
\begin{equation}
G(x,x')=\frac{1}{2}\left[ A(x,x')+C(x,x')\right] =
\int \frac{d^4 k}{(2\pi)^4}e^{ik(x-x')}\frac{i\hbar }{k^2+i \eta }=
\int \frac{d \omega \,d^3k}{(2\pi)^4}
\frac{ic\hbar}
{ \omega ^2- \Omega _{\bf k}^2+i \eta  } e^{-i[ \omega (t-t')-
{\bf k}({\bf x}-{\bf x}')]},
\label{@}\end{equation}
where $ \eta $ is an infinitesimally small number $>0$.

We shall now focus attention
upon nonrelativistic systems which are so small that
the effects of retardation can be
neglected.
Then we can ignore the ${\bf x}$-dependence
in (\ref{18.386a}) and (\ref{18.386})
and find
\begin{equation}
C^{ij}_{ \rm b}({\bf x}\,t,{\bf x}'\,t')
\approx
C^{ij}_{ \rm b}(t,t')=-i \frac{\hbar }{6\pi^2}\delta ^{ij} \partial _t\delta ^R(t-t').
\label{@GREE}\end{equation}
The superscript $R$ indicate that the
$ \delta $-function is a right-sided one,
with the property
\begin{equation}
\int dt\, \Theta(t)  \delta ^R(t)=1.
\label{@}\end{equation}
 The retarded nature of the $ \delta $-function
expresses the \iem{causality}
of the dissipation forces, which
is
crucial for producing a
probability conserving
time evolution of the probability
distribution \cite{PITOPF}.

 Inserting this into (\ref{18.148bt}) and integrating by parts,
we obtain two contributions.
The first is a diverging term
\begin{eqnarray}
 \Delta {\cal A}_{\rm loc}[{\bf x}_+,{\bf x}_-]=
\frac{ \Delta M}{2} \int_{t_a}^{t_b} dt \,
  (\dot {\bf x}_+^2-\dot {\bf x}_-^2)(t)
,
\label{18.inflfC3}
\end{eqnarray}
 where
\begin{equation}
\!\!\!\!\!\!\!\!\! \Delta {M}\equiv-\frac{e^2}{c^2}
\int \frac{d\omega'd^3k}{(2\pi)^4}
\frac{\sigma _{\bf k} (\omega ')}{\omega '}
\delta ^{ij\,\rm tr}_{{\bf kk}}
=-\frac{e^2}{6\pi^2c^3} \int_{-\infty}^\infty dk.
\label{18.deltaome}\end{equation}
 diverges linearly. This simply renormalizes
the  kinetic terms in the path integral
(\ref{18.389n}), renormalizing them to
\begin{eqnarray}
 \frac{i}{\hbar}\int\tab dt\,
    \frac{M_{\rm ren}}2\left(\dot {\bf x}_+^2-\dot  {\bf x}_-^2\right).
 \label{18.389nn}
\end{eqnarray}
By
 identifying $M$ with $M_{\rm ren}$ this renormalization may be ignored.

The second term has the form
\begin{eqnarray}
\!\!\!\!\!
\!\!\!\!\!\!\!\!\!\!\!\!\!
{\cal A}_D^{\rm FV}[{\bf x}_+,{\bf x}_-]&=&  -\gamma
\frac{M}{2}\int_{t_a}^{t_b} dt\,
(\dot {\bf x}_+-\dot {\bf x}_-)(t)
(\ddot {\bf x}_++\ddot {\bf x}_-)^R(t),
\label{18.inflfC}\end{eqnarray}
with the friction constant
\begin{equation}
 \gamma \equiv \frac{e^2}{6\pi c^3M}=\frac{2}{3}\frac{ \alpha }{ \omega _M},
\label{18.gammafu2}\end{equation}
where $
 \alpha\equiv{e^2}/{\hbar c}\approx{1}/{137}$ is the feinstructure constant
and $\omega _M \equiv {Mc^2}/{\hbar }$ the Compton frequency
associated with the mass $M$.
 In contrast to the ordinary friction
constant, this has the dimension $1/$frequency.
A dependence of the mass renormalization on the potential $V$
(the Lamb shift) will be discussed in
Sec.\ IV.

The superscript $R$ in (\ref{18.inflfC}) indicates that the
the
 acceleration
 $(\ddot {\bf x}_++\ddot {\bf x}_-)(t)$
is slightly shifted with respect to the velocity factor $(\dot {\bf x}_+-\dot  {\bf x}_-)(t)$
towards an earlier time.
This accounts for the fact that
a physical friction force
is always retarded, due to the Heaviside
function in (\ref{18.148bt}).
 The retardation
expresses the \iem{causality}
of the dissipation forces, which
is
crucial for producing a
probability conserving
time evolution of the probability
distribution \cite{PITOPF}.

We now turn to the anticommutator function. Inserting
(\ref{18.387}) and the friction constant $ \gamma $ from (\ref{18.gammafu2}),
it becomes
\begin{eqnarray}
\frac{e^2}{c^2}A_{\rm b}({\bf x}\,t,{\bf x}'\,t' )
\approx 2 \gamma k_BTK(t,t')
 ,   \label{18.386k}
\end{eqnarray}
where
\begin{eqnarray}
K(t,t')=K(t-t')&\equiv&
  \int _{-\infty}^\infty \frac{d \omega'}{2\pi}
K( \omega')
e^{-i \omega'(t-t')}
 ,   \label{18.390}
\end{eqnarray}
with a Fourier transform
\begin{eqnarray}
K( \omega')\equiv
\frac{\hbar  \omega' }{2k_BT} \coth\frac{\hbar  \omega' }{2k_BT},
    \label{18.386k3}
\end{eqnarray}
whose high-temperature expansion starts out like
\begin{eqnarray}
K( \omega')\approx K^{\rm HT}( \omega')\equiv1-\frac{1}{3}\left(\frac{\hbar  \omega' }{2k_BT}\right)^2.
    \label{18.391}
\end{eqnarray}
The function $K( \omega')$
has
the
normalization $K(0)=1$,
giving $K(t-t')$
a unit temporal area:
\begin{equation}
\int_{-\infty}^{\infty} dt\, K(t-t') = 1.
  \label{18.392}
\end{equation}
Thus $K(t-t')$ may be viewed as a $ \delta$-function broadened
by quantum  fluctuations and relaxation effects.

With the function $K(t,t')$, the fluctuation part of the influence functional
in
(\ref{18.148btA}),
(\ref{18.148bt}),
(\ref{18.389n}) becomes
\begin{equation}
{\cal A}_F^{\rm FV}[{\bf x}_+,{\bf x}_-]=
i\frac{w}{2\hbar}
    \int_{t_a}^{t_b} dt\, \int_{t_a}^{t_b} dt'\,
    (\dot {\bf x}_+ -\dot {\bf x}_-)(t)\,K(t,t')\,(\dot {\bf x}_+ -\dot {\bf x}_-)(t').
\label{18.inflfA}\end{equation}
Here we have used the symmetry of the function
$K(t,t')$ to remove the Heaviside function
$\Theta(t-t')$ from the integrand, extending the range of $t'$-integration
to the entire interval $(t_a,t_b)$.
We have also introduced the constant
\begin{equation}
w\equiv
2M\gamma k_B T,
\label{@wconst}\end{equation}
for brevity.

At very
 high temperatures, the time evolution amplitude for
the density matrix
 is given by the path integral
\begin{eqnarray}
\!\!U({\bf x}_{+b},{\bf x}_{-b},t_b|
{\bf x}_{+a},{\bf x}_{-a},t_a
) &=&
  \int {\cal D}{\bf x}_+(t)\int {\cal D}{\bf x}_-(t)\,
    \exp\left\{
 \frac{i}{\hbar}\int\tab dt\,
    \left[ \frac{M}{2}({\dot {\bf x}_+}^2-{\dot {\bf x}_-}^2)-
    (V({\bf x}_+)-V({\bf x}_-))\right]
    \right\} \nonumber \\
&\times&
\exp\left\{ -\frac{i}{2\hbar }M \gamma \int\tab dt\,
\,(\dot  {\bf x}_+-\dot {\bf x}_-)
(\ddot {\bf x}_++\ddot {\bf x}_-)^R
-
   \frac{w}{2\hbar^2}
    \int_{t_a}^{t_b} dt\,
    (\dot {\bf x}_+ -\dot {\bf x}_-)^2
  \right\},
\label{18.389}
\end{eqnarray}
where
the last term becomes local
for high temperatures,
since $K(t,t')\rightarrow  \delta (t-t')$.
This
is the \iem{closed-time path integral\/}
of a particle in contact with a thermal reservoir.
For moderately high temperature, we should include
also the first correction term
in (\ref{18.391}) which adds to the exponent an additional term.
\begin{equation}
   \frac{w}{24 (k_BT)^2}
    \int_{t_a}^{t_b} dt\,
    (\ddot {\bf x}_+ -\ddot {\bf x}_-)^2.
    \label{@Dterm}\end{equation}

In the classical limit, the last
squeeses the forward and backward
paths together. The density matrix
(\ref{18.389}) becomes diagonal. The $ \gamma $-term, however,
remains and describes classical radiation damping.

\comment{It is now convenient to change integration variables
and go over to average and relative
coordinates of the two paths ${\bf x}_+$, ${\bf x}_-$:
\begin{eqnarray}
{\bf x} & \equiv  & ({\bf x}_ + +{\bf x}_-)/2, \nonumber \\
{\bf y}&\equiv & {\bf x}_+ -{\bf x}_-.
  \label{18.393}
\end{eqnarray}
Then (\ref{18.389}) becomes
\begin{eqnarray}
U({\bf x}_{+b},{\bf x}_{-b},t_b|
{\bf x}_{+a},{\bf x}_{-a},t_a
) &=&
  \int {\cal D}{\bf x}(t)\int  {\cal D}{\bf y}(t)\,
 \exp\bigg\{
-\frac{i}{\hbar}
    \int \tab dt\, \left[M \left(-\dot {\bf y}\dot {\bf x}+
\gamma \dot {\bf y}\ddot {\bf x}^R\right)+
V\left({\bf x}+\frac{{\bf y}}{2}\right)-V\left({\bf x}-\frac{{\bf y}}{2}\right)\right]
    \nonumber \\
&   &\!\,\!\!\!\!\!\!\!\!~~~~~~~~~~~~~~~~~~~~~~~~~~~~~~~~~~~- \,
     \frac{w}{2\hbar^2}
    \int\tab dt\, \int_{t_a}^{t_b} dt'\,
      \dot {\bf y}(t)K(t,t')\dot {\bf y}(t') \bigg\}.
  \label{18.394}
\end{eqnarray}
}

\section{Langevin Equations}

For
 high $ \gamma T$,
the last term in the forward--backward path integral
(\ref{18.389})
makes  the
size of the fluctuations in the difference between the paths
 ${{\bf y}}(t)
\equiv
{\bf x}_+(t)-
{\bf x}_-(t)$ very small.
It is then convenient to  introduce the average of the two
paths as
 ${{\bf x}}(t)
\equiv   \left[
{\bf x}_+(t)+{\bf x}_-(t)\right] /2$,
and expand
\begin{equation}
V\left({\bf x}+\frac{{\bf y}}{2}\right)-V\left({\bf x}-\frac{{\bf y}}{2}\right) \sim
  {\bf y}\cdot{\BF\nabla}V({\bf x})+{\cal O}({\bf y}^3)\dots~,
  \label{18.395}
\end{equation}
keeping only the first term.
We
further
introduce an auxiliary quantity ${\BF \eta} (t)$ by
\begin{equation}
\dot{\BF \eta} (t)\equiv M\ddot {\bf x}(t)- M\gamma\!\!\dddot{{\bf x}}(t)+\BF \nabla V({\bf x}(t)).
  \label{18.396}
\end{equation}
With this, the
exponential function in (\ref{18.389}) becomes
\begin{equation}
\exp\left[- \frac{i}{\hbar}
  \int_{t_a}^{t_b}
   dt\, \dot {\bf y}  {\BF\eta} - \frac{w}{2\hbar^2}
\int_{t_a}^{t_b}  dt\,
    \dot {\bf y}^2(t) \right],
  \label{18.397}
\end{equation}
where
  $ w $ is the constant (\ref{@wconst}).

Consider now the diagonal part of the amplitude
(\ref{18.395}) with
${\bf x}_{+b}=
{\bf x}_{-b}\equiv {\bf x}_b$
and
${\bf x}_{+a}=
{\bf x}_{-a}\equiv {\bf x}_a$, implying that
${\bf y}_b={\bf y}_a=0$.
It represents a probability distribution
\begin{equation}
P({\bf x}_b\,t_b|{\bf x}_a\,t_a)\equiv |( {\bf x}_{b},t_b|{\bf x}_{a},t_a)|^2
\equiv
U({\bf x}_{b},{\bf x}_{b},t_b|
{\bf x}_{a},{\bf x}_{a},t_a
).
\label{@18.probl1}\end{equation}
Now
the
 variable ${\bf y}$ can simply be integrated out
in (\ref{18.397}),
 and we find
the probability distribution
\begin{equation}
P[ \BF\eta ]\propto \exp\left[ -\frac{1}{2w}
  \int_{t_a}^{t_b}  dt\,
    {\BF \eta}^2(t)
  \right].
  \label{18.398}
\end{equation}
The expectation value of an arbitrary
functional of $F[x]$
can be calculated from the path integral
\begin{eqnarray}
\langle F[{\bf x}]\rangle_ \eta  \equiv {\cal N}
\int {\cal D}{\bf x}\,  P[\BF\eta ]F[{\bf x}],
\label{@Expecx}\end{eqnarray}
where the normalization factor ${\cal N}$ is fixed by the condition
$\langle \,1\,\rangle=1$.
By a change of integration variables from $x(t)$ to $ \eta (t)$,
the expectation value (\ref{@Expecx})  can be rewritten as a
functional integral
\begin{eqnarray}
\langle F[{\bf x}]\rangle _ \eta \equiv {\cal N}
\int {\cal D} \BF\eta \,  P[ \BF\eta ]\, \,F[{\bf x}],
\label{@Expece}\end{eqnarray}
Note that the probability distribution (\ref{18.398})
is $\hbar$-independent. Hence in the approximation
(\ref{18.395}) we obtain the classical Langevin equation.
In principle, the integrand contains
a factor $J^{-1}[x]$, where $J[x]$ is the functional Jacobian
\begin{equation} \label{18.deter}
J[{\bf x}]\equiv\Det[\lfrac{  \delta  \eta^i (t) }{ \delta x^j(t')}]=
 \det[\left(M\partial _t^2 - M\gamma \partial_t^{3\hspace{.2pt}R}\right) \delta _{ij}
+\nabla_i\nabla_j V({\bf x}(t))].
 \label{18.fjac}
\end{equation}
It can be shown that
the determinant is
unity, due to the retardation
of the friction term \cite{PITOPF}, thus justifying its omission
in (\ref{@Expece}).

The path integral
(\ref{@Expece})
may be interpreted as
an expectation value with respect to the solutions
of a \iem{stochastic differential equation}
(\ref{18.396}) driven by a
 Gaussian
random \iem{noise} variable $\eta(t) $
with a
correlation function
\begin{equation}
\langle \eta^i(t) \eta^j(t') \rangle_T = \delta ^{ij}
w \,  \delta (t-t').
  \label{18.399}
\end{equation}
Since the dissipation carries a third time derivative, the
treatment of the initial conditions is nontrivial
and will be
discussed elsewhere.
In most physical applications $ \gamma $ leads to slow decay rates.
In this case
the
simplest procedure to solve
(\ref{18.396}) is
to write
the
stochastic equation as
\begin{equation}
 M\ddot {\bf x}(t)
+\BF \nabla V({\bf x}(t))=
\dot{\BF \eta} (t)
+
 M\gamma\!\!\dddot{{\bf x}}(t),
  \label{18.396n}
\end{equation}
and solve it iteratively,
first without the $ \gamma$-term, inserting
the solution  on the right-hand side,
and  such a procedure is equivalent to a perturbative expansion
in $ \gamma $ in Eq.~(\ref{18.389}).

\section{Master Equation for Time Evolution
of Density Matrix}

We now derive a
Schr\"odinger-like differential equation
describing the evolution of the density
matrix $ \rho (x_{+a},x_{-a};t_a)$
in Eq.~(\ref{@locev}).
In the standard  derivation of such an equation \cite{PI}
one first localizes the
last term via a quadratic completion involving a
fluctuating noise variable $ \eta (t)$.
 Then one goes over to a canonical
formulation of the path integral
(\ref{18.389}), by
rewriting it as
a path integral
\begin{eqnarray}
\!\!\!\!\!\!\!\!\!\!\!\!\!\!\!\!\!\!\!\!
U_ \eta ({\bf x}_{+b},{\bf x}_{-b},t_b|
{\bf x}_{+a},{\bf x}_{-a},t_a
) &=&
  \int {\cal D}{\bf x}_+(t)\int {\cal D}{\bf x}_-(t)
  \int \frac{{\cal D}{\bf p}_+}{(2\pi)^3}
  \int \frac{{\cal D}{\bf p}_-}{(2\pi)^3}
  \nonumber \\
\!\!\!\!\!\!\!\!\!\!\!\!\!\!\!\!\!\!\!\!\! &\times&     \exp\left\{
 \frac{i}{\hbar}\int\tab dt\,
    \left[
{\bf p}_+\dot{\bf x}_+
-{\bf p}_-\dot{\bf x}_-
 -{\cal H}_  \eta  ({\bf p}_+,{\bf p}_-,{\bf x}_+ ,{\bf x}_-)
\right] \right\} .
 \label{X18.389nn}
\end{eqnarray}
Then $U_ \eta ({\bf x}_{+b},{\bf x}_{-b},t_b|
{\bf x}_{+a},{\bf x}_{-a},t_a
)$ satisfies the differential equation
\begin{equation}
i\hbar\partial _t
 U_ \eta (x\,,y\,,t|x_a,y_a,t_a)
=\hat{\cal H}_\eta  \,U_ \eta (x\,,y\,,t|x_a,y_a,t_a).
\label{18.sla2}\end{equation}
The same equation is obeyed by the density matrix
$ \rho (x_{+},x_{-};t_a)$.

At high temperatures
where the action
in the path integral (\ref{18.389}) is local,
we can find directly a Hamiltonian  without the  noise-averaging procedure.
However, the standard procedure
of finding a canonical formulation
is not applicable
because of the high time derivatives of ${\bf x}(t)$
in the action
of (\ref{18.389}). They
can be transformed into canonical momentum variables
only by introducing
several auxiliary independent
variables
${\bf v}\equiv  \dot {\bf x}$,
${\bf b}\equiv  \ddot {\bf x}, \dots$~\cite{ddot,GFCM}.
For small dissipation, which we shall consider,
it is preferable to
proceed in another way
by going first to a canonical formulation
of the  quantum system without the
effect of electromagnetism, and include the effect
of the latter
recursively.
For simplicity, we shall treat only the local limiting form of the last term in
(\ref{18.389}).
In this limit, we define
a Hamilton-like operator as follows:
\begin{equation}
\!\!\!\!\hat{\cal H}
\equiv \frac{1}{2M}\left(\hat {\bf p}_+^2-\hat {\bf p}_-^2\right)
+V({\bf x}_+)-V({\bf x}_-)
+\frac{M \gamma }{2}
\,(\hat{\dot{\bf x}}_+-\hat{\dot {\bf x}}_-)
(\hat{\ddot {\bf x}}_++\hat{\ddot {\bf x}}_-)^R
     -i
   \frac{w}{2\hbar}
    (\hat{\dot {\bf x}}_+ -\hat{\dot {\bf x}}_-)^2
.
\label{canhn2p}\end{equation}
Here $
\hat{\dot {\bf x}}$,
$\hat{\ddot {\bf x}}$
are abbreviations for the commutators
\begin{equation}
\hat{\dot {\bf x}}\equiv \frac{i}{\hbar }[\hat{\cal H},\hat{{\bf x}}]
 ,
~~~~
\hat{\ddot {\bf x}}\equiv \frac{i}{\hbar }[\hat{\cal H},\hat{\dot{\bf x}}].
\label{@}\end{equation}
A direct differentiation of Eq. (\ref{18.389}) over time leads to
the conclusion that
 the density matrix
$
 \rho (x_{+},x_{-};t_a)
$ satisfies the time evolution
equation
\begin{equation}
i\hbar \partial _t
 \rho (x_{+},x_{-};t_a)=
\hat{\cal H}
 \rho (x_{+},x_{-};t_a).
\label{@rhoe}\end{equation}

At moderately high temperatures, we also include a term
coming from (\ref{@Dterm})
\begin{equation}
    {\cal H}_1\equiv
i
   \frac{w}{24 (k_BT)^2}
    (\hat{\ddot {\bf x}}_+ -\hat{\ddot {\bf x}}_-)^2.
    \label{@3.6}\end{equation}
For systems with fricition caused by a conventional heat bath
of harmonic oscillators as discussed by Caldeira and Leggett
\cite{CL}, the analogous extra term
was shown by Diosi \cite{Dio}
to bring the Master equation to the general
Lindblad form \cite{LB}
which ensures positivity of the probabilities
resulting from the solutions
of (\ref{@rhoe}).

It is useful to re-express (\ref{@rhoe}) in
the standard quantum-mechanical operator form
where the density matrix has a
bra--ket representation
$\hat  \rho (t)=\sum _{mn} \rho_{nm} (t)|m\rangle\langle n|$.
Let us denote the initial Hamilton operator
of the system in (\ref{@PRO}) by $\hat H = {\hat {\bf  p}^2}/{2M} + \hat V$,
then
Eq.~(\ref{@rhoe})  with the term  (\ref{@3.6})
takes the operator form
\begin{eqnarray}
i\hbar  \partial_t \hat\rho & = & \hat {\cal H}\,\hat  \rho
 \equiv [\hat H, \hat  \rho ]
 + \frac{M\gamma }{2}\left(
\hat{\dot {\bf x}}
\hat{\ddot {\bf x}}\hat\rho
+\hat \rho
\hat{\ddot {\bf x}}\hat{\dot{\bf x}}
-\hat{\dot {\bf x}}\,\hat\rho\, \hat{\ddot {\bf x}}
-\hat{\ddot {\bf x}}\,\hat\rho\, \hat{\dot {\bf x}}
\right)
- \frac{i w}{2} [
\hat{\dot {\bf x}},[
\hat{\dot {\bf x}},
\hat   \rho ]]
+ \frac{i w\hbar ^2}{24(k_BT)^2} [
\hat{\ddot {\bf x}},[
\hat{\ddot {\bf x}},
\hat   \rho ]] .{}
\label{3.5}\end{eqnarray}
 The retardation of $\ddot{\bf x}_\pm$
in (\ref{canhn2p}) leads to the specific operator order
in the second term,
which ensures that
Eqs.\ (\ref{canhn2p})
and  (\ref{3.5}) preserve the total probability.

For a free particle
with $ V = 0$
and
  $ [{\cal H }, {\bf p}] = 0$, one has $\hat{\dot{\bf x}}_\pm
  = \hat {\bf p}_\pm /M$
to all orders
 in $ \gamma $,
such that the
time evolution
equation (\ref{3.5})
becomes
\begin{eqnarray}
i\hbar  \partial_t \hat\rho & = &
  [\hat H, \hat  \rho ]
- \frac{i w}{2M^2} [\hat{\bf p},[
\hat{{\bf p}},
\hat   \rho ]] .{}
\label{3.5s}\end{eqnarray}
In the momentum representation of the density matrix
 $\hat  \rho =\sum _{{\sbf p}{\bf p}'} \rho _{{{\sbf p}{\bf p}'}}|{\bf p}\rangle\langle {\bf p}'|$,
  the last term simplifies to
 $-i \Gamma \equiv - {i w}({\bf p-{\bf p}')^2}/{2M^2}$
showing that
a
free particle does not dissipate
energy by radiation, and that the
off-diagonal matrix elements decay with the
  rate $  \Gamma $.

In general, Eq.~(\ref{canhn2p})
is  an implicit equation for
the Hamilton operator
$\hat{\cal H}
$. For small $e^2$ it can be solved approxiamtely in a single
iteration step, replacing
$\hat{\dot {\bf x}}$ by ${\hat{\bf p}}/M$ and
${\hat{\ddot {\bf x}}}=-{\BF \nabla} V/M$ in Eq.~(\ref{3.5}).

The validity of  this iterative procedure is most easily proven in the
time-sliced path integral. The final slice
of infinitesimal width $ \epsilon $
reads
\begin{eqnarray}
\!\!\!\!\!\!\!\!\!\!\!\!\!\!\!\!\!\!\!\!
U ({\bf x}_{+b},{\bf x}_{-b},t_b|
{\bf x}_{+a},{\bf x}_{-a},t_b- \epsilon
) &=&
  \int \frac{{d}{\bf p}_+(t_b)}{(2\pi)^3}
  \int \frac{{d}{\bf p}_-(t_b)}{(2\pi)^3}
e^{
 \frac{i}{\hbar}
    \left\{
{\bf p}_+(t_b)
\left[ {\bf x}_+(t_b)-
{\bf x}_+(t_b- \epsilon )\right]
-{\bf p}_-\dot{\bf x}_-
 -{\cal H}(t_b)
\right\}}
 .
 \label{x18.389nn}
\end{eqnarray}
Consider now a term of the generic form
$\dot F_+({\bf x}_+)
F_-({\bf x}_-)$ in ${\cal H}$.
When differentiating $U ({\bf x}_{+b},{\bf x}_{-b},t_b|
{\bf x}_{+a},{\bf x}_{-a},t_b- \epsilon)$
with respect to the final time $t_b$,
the integrand receives a factor
$-{\cal H}(t_b)$.
The term
$\dot F_+({\bf x}_+)
 F_-({\bf x}_-)$ in ${\cal H}$
has the explicit form
$ \epsilon ^{-1}\left[  F_+({\bf x}_+(t_b))
-  F_+({\bf x}_+(t_b- \epsilon ))\right]
  F_-({\bf x}_-(t_b))
$.     It it can be taken out of the integral, yielding
\begin{equation}
 \epsilon ^{-1}\left[  F_+({\bf x}_+(t_b))U
-U  F_+({\bf x}_+(t_b- \epsilon ))\right]
  F_-({\bf x}_-(t_b)).
\label{@}\end{equation}
In operator language, the amplitude $U$ is equal to
$\hat U\approx 1-i \epsilon {\hat{\cal H}}/\hbar$, such the term
$\dot F_+({\bf x}_+)
F_-({\bf x}_-)$ in ${\cal H}$
yields an
operator
\begin{equation}
 \frac{i}{\hbar }
\left[\hat{\cal H}, \hat F_+({\bf x}_+)
\right] F_-(x_-)
\label{@}\end{equation}
in the differential operator for the time evolution.

For functions of the second derivative $\ddot{\bf x}
$  we have
to split off the last two time slices
and convert the two intermediate integrals over ${\bf x}$
into
 operator expressions, which obviously leads to the repeated commutator
of $\hat {\cal H}$ with $\hat{\bf x}$, and so on.
\section{Line Width}
Let us apply the master equation (\ref{3.5})
to atoms, where $V({\bf x})$ is the Coulomb potential,
 assuming it to be initially in an eigenstate $|i\rangle$
 of $ H$, with a density matrix
 $\rho = |i  \rangle \langle i|$.
 Since atoms decay rather slowly, we may treat the $ \gamma $-term in (\ref{3.5})
perturbatively. It leads to a time derivative
of the density matrix
\begin{eqnarray}
 \partial_t \langle i |  \hat\rho (t) | i \rangle & = &- \frac{  \gamma }{ \hbar M}
		  \langle i | [ \hat H, \hat {{\bf p}}] \,\hat{\bf p}\,\hat  \rho(0) | i \rangle
  =    \frac{\gamma}{M}  \sum_{f\neq i}  \omega _{fi} \langle i | {\bf p} | f\rangle\langle f|
   {\bf p} | i\rangle
 =   - M  \gamma  \sum_f  \omega _{fi}^3 \, |{\bf x}_{fi}|^2.
\label{3.8n}\end{eqnarray}
where $ \hbar \omega _{fi} \equiv E_i -  E _f $ and $
{\bf x}_{fi}\equiv   \langle f | {\bf x} | i\rangle
$ are the matrix elements
of the dipole operator.
A further contribution comes the last two terms in
(\ref{3.5}):
\begin{eqnarray}
\partial_t \langle i|  \rho |i\rangle = -\frac{ w }{M^2 \hbar^2 }
\langle i| {\bf p}^2 |i\rangle
+
\frac{ w }{12 M^2 (k_BT)^2 }
\langle i| \dot{\bf p}^2 |i\rangle
  =
- w    \sum_{n}   \omega ^2_{fi}\left[1-\frac{\hbar ^2 \omega ^2_{fi}}{12 (k_BT)^2 }\right]\, | {\bf x}_{fi}|^2.
\label{3.9n}\end{eqnarray}
This time dependence
is caused by spontaneous emission and induced
emission and absorption.
To identify these terms
we rewrite the spectral content
of a single oscillator of frequency $ \Omega$
in the local approximation
$A^{ij}_{ \rm b}(t,t')  + C^{ij}_{ \rm b}(t,t')$
to the correlation functions in (\ref{18.386a})
and (\ref{18.386}) in the ${\bf x}$-independent approximation
as
\begin{equation}
\!\!\!\!\!\!\!\!
\displaystyle
 C_b(t,t')+
A_b(t,t')=\frac{4\pi}{3}\pi\hbar \int \frac{d \omega d^3k}{(2\pi)^4}
 \frac{\pi }
{2M{\Omega_{\sbf k } }}
  \left(\coth \frac{\omega'  }{2k_BT} +1
\right)
  \left[\delta(\omega ' -\Omega_{\sbf k }  )-\delta(\omega ' +\Omega_{\sbf k }  )\right]e^{-i \omega' (t-t')} ,
  \label{18.155}
\end{equation}
as
\begin{equation}
~~\!\!\!\!\!\!\!\!   \displaystyle
 C_b(t,t')+
A_b(t,t')=\frac{4\pi}{3}\hbar \int \frac{d \omega d^3k}{(2\pi)^4}
\frac{\pi }{M{\Omega }}
\left\{  \delta (\omega' -\Omega_{\sbf k }  )+\frac{1}{e^{\Omega_{\sbf k }  /k_BT}-1}
[\delta (\omega' -\Omega_{\sbf k } )
+\delta (\omega' +\Omega_{\sbf k } )]
\right\}e^{ -i \omega' (t-t')}.
  \label{18.156b}
\end{equation}
Following Einstein's intuitive interpretation,
the first term in curly brackets is due to spontaneous emission,
the other two terms accompanied by
the Bose
occupation function
account for induced emission and absorption.
In the large-$T$
expressions
(\ref{3.8n}) and
(\ref{3.9n}), the first term in  (\ref{18.156b})
leads to part of the sum
with $ \omega _{fi}>0$ only.
This is the famous Wigner-Weisskopf
formula for the natural line width of atomic levels.

The rest of the time dependence
is due to induced
absorption and emission.

\section{Lamb shift}
For atoms, the calculation of the diverging term
(\ref{18.inflfC3}) can be done in a more sensitive approximation.
Being interested in the time behavior
of the pure-state density matrix
$\rho = |i  \rangle \langle i|$, we may calculate the affect
of the action (\ref{18.148bt}) perturbatively
as follows:
Consider the first term in the local approximation,
and integrate the external positions in the path integral
(\ref{18.389n})
over the initial wave functions, forming
\begin{equation}
 U_{ii,t_b;ii,t_a}=
\int d{\bf x}_{+b}\,d {\bf x}_{-b}
\int d{\bf x}_{+a}\,d {\bf x}_{-a}
\langle i|{\bf x}_{+b}\rangle
\langle i|{\bf x}_{-b}\rangle
U({\bf x}_{+b},{\bf x}_{-b},t_b|
{\bf x}_{+a},{\bf x}_{-a},t_a
)
\langle {\bf x}_{+b}|i\rangle
\langle {\bf x}_{-b}|i\rangle
\label{@}\end{equation}
To lowest order in $ \gamma $,
the contribution of
(\ref{18.148bt}) can be evaluated as follows
\begin{eqnarray}
 \Delta U_{ii,t_b;ii,t_a}&\!\!=&i\frac{e^2}{\hbar ^2c^2}
\int_{t_a}^{t_b}\!\! dtdt'  \sum_f
\int d{\bf x}_{+}\!
\int d{\bf x}'_{+}\,
U_{ii,t_a;ii,t}
\langle i|{\bf x}_{+}\rangle
{\bf x}_+
\langle {\bf x}_{+}|f\rangle
\partial_t\partial _{t'}  C^{ij}_{ \rm b}(t,t')
U_{fi,t;fi,t'}
\langle f|{\bf x}'_{+}\rangle
{\bf x}'_{+}
\langle {\bf x}'_{+}|i\rangle
U_{ii,t';ii,t_a}   \nonumber \\&&
\label{@}\end{eqnarray}
Inserting $U_{ii,t_a;ii,t}=e^{-iE_i(t_b-t)}$ etc.,
 this becomes
\begin{equation}
\Delta U_{ii,t_b;ii,t_a}=i\frac{e^2}{\hbar ^2c^2}
\int_{t_a}^{t_b} dt dt' \, \partial_t\partial _{t'}  C_b(t,t')
\langle i | \hat{ {\bf x}} (t) \hat{{{\bf x}}'}(t') | i \rangle =
i\frac{e^2}{\hbar ^2c^2}
\sum_f\int_{t_a}^{t_b} dt dt' \,  C_b(t,t')
\langle i | \hat{\dot {\bf x}} (t)
|f\rangle
\langle f|
 \hat{\dot {{\bf x}}'}(t') | i \rangle .
\label{@}\end{equation}
The integration over $t$ and $t'$
yields
\begin{eqnarray}
 \Delta U_{ii,t_b;ii,t_a}&\!\!=&i\frac{e^2}{\hbar ^2c^2}\int_{t_a}^{t_b} dt \int  \frac{d \omega ' d^3 k}{( 2\pi )^4}
 \sigma _k (     \omega ') \sum_{f} \frac{1}{E_{i}- E_f -  \omega '
 -i\eta} |   \hat{\dot{\bf x}}_{fi}  |^2
\label{4.1}\end{eqnarray}
After subtracting the mass renormalization
(\ref{18.inflfC3}),
which can be rewritten
once more in the same form as in (\ref{4.1})
with $E_i-E_f=0$,
the integral yields
$\ln [(  \Lambda  + E_i - E_f)/|E_i-E_f|]$
where $ \Lambda $ is Bethe's cutoff \cite{Bethe}.
For $ \Lambda \gg |E_i - E_f |$, the result (\ref{4.1})
imples an energy shift of the
atomic level $|i\rangle$:
\begin{equation}
 \Delta E_i=\frac{e^2}{\hbar c^3}\frac{2\hbar  }{3\pi}
\sum _f  \omega ^3_{fi}  |   \hat{{\bf x}}_{fi}  |^2\log\frac{ \Lambda }{ |\omega _{fi}|},
\label{@}\end{equation}
which
is the celebrated Lamb shift.
Usually, the logarithm is approximated by a weighted
 average $L$ over energy
levels.
Then
contribution of
 the term (\ref{4.1})
can be attributed to an extra
term in the Hamiltonian (\ref{3.5}) $H_{\rm LS}=iL M  \gamma /2
 [\hat {\dot x}, , \hat {\ddot x}]$.

\section{Conclusion}
We have calculated the master equation for the time evolution of the
quantum mechanical density matrix
describing dissipation and decoherence
of a point particle interacting with the electromagnetic field.
The Hamilton-like evolution operator
was specified recursively.
To lowest order in the electromagnetic coupling strength,
we have recovered the known Lamb shift and natural line width of
atomic levels. In addition, we have calculated the
additional broadening caused
by the coupling of the photons to the thermal bath.

 This research is supported by German HSP III grant.

\end{document}